
\input phyzzx
\PHYSREV
\hoffset=0.3in
\voffset=-1pt
\baselineskip = 14pt \lineskiplimit = 1pt
\frontpagetrue
\rightline {Cincinnati preprint Dec-1993}
\medskip
\titlestyle{\seventeenrm Phase Transition and Absence
            Of Ghosts in Rigid QED}
\vskip.5in
\medskip
\centerline {\caps M. Awada and D.Zoller\footnote*{\rm E-Mail
             address: moustafa@physunc.phy.uc.edu}}
\centerline {Physics Department}
\centerline {\it University of Cincinnati,Cincinnati, OH-45221}
\bigskip
\centerline {\bf Abstract}

Ordinary QED formulated in the Feynman's space-time picture is
equivalent to a one dimensional field theory.  In the large N limit
there is no phase transition in such a theory.  In this letter, we show
a phase transition does exist in a
generalization of QED characterized by the addition of the curvature of
the world line (rigidity) to the Feynman's space-time action.  The large
distance scale of the disordered phase essentially coincides with
ordinary QED, while the ordered phase is strongly coupled.  Although
rigid QED exhibits the typical pathologies of higher derivative theories
at the classical level, we show that both phases of the quantum theory
are free of ghosts and tachyons.  Quantum fluctuations prevent taking
the naive classical limit and inherting the problems of the classical
theory.
\eject

{\bf I-The Rigid Model Of QED}

The second quantized continuum theory of QED is essentially uniquely
determined by the demand of renormalizability. A naive generalization
would therefore be difficult.   The current theory seems to exist in one
phase, the weak coupling phase, where the Coulomb coupling constant
increases indefinetly at small distance scales indicating a breakdown of
the theory at the Landau singularity.  The quest of a  strong phase of
continuum QED remains therefore an open question.  The first quantized
description of the Feynman's space-time formulation, however, admits the
renormalizable generalization of adding the scale invariant curvature of
the world line (rigidity) to the usual relativistic point
particle-Maxwell action where the matter sector is described by
$x^{\mu}, \mu=1,2,...D$ and the gauge sector by the usual Maxwell U(1)
gauge field $A_{\mu}$.  In [1], [2] we have shown that the curvature
term arises out of a loop splitting regularization of electrodynamics
and conjectured the existence of a phase transition due to the long
range Coulomb interactions.  Here we prove this conjecture and also show
 this higher derivative theory is free of ghosts and tachyons that
typically plague higher derivative theories.

The conjecture that there is a phase transition can be easily understood
by analogy with spin systems.  The arc-length gauge fixing condition
[3] of the relativistic point particle resembles the sigma model
constraint $\sigma^{2} =1$ where the velocity of the particle
corresponds to the direction of the spin.  Our action resembles
a modified sigma model in one dimension with long range interactions
[2] where the curvature term plays the role of the kinetic term of
the spin field.  Since spin systems with sufficiently longe range
interactions may exhibit a phase transition in one dimension, we
conjecture likewise for rigid QED.  The large N is a successfull
approximation for non-linear sigma models and we applied it to
rigid QED where N is the space-time dimension D. We found a critical
line in the plane of the Coulomb coupling verses the curvature
coupling below which there is a disordered phase and above which is
a new ordered strongly coupled phase.

The higher derivative nature of QED should cause a serious pause as
any higher derivative theory is typically pathological. Indeed,
the arc-length plus the curvature term theory has classical runaway
solutions which are tachyonic.  Whether a higher derivative regulated
quatum theory has such  pathological behaviour is more subtle and
depends on details  of the continuum limit.  A free scalar field
theory on the lattice with spacing ${1\over \Lambda}$, has higher
derivatives and ghosts.  However these ghosts have mass of order
$\Lambda$ and decouple in the continuum limit as
$\Lambda\rightarrow\infty$.  In the less trivial case of rigid QED
we will show that the ghosts have mass of order $\Lambda$ and similarly
decouple from the continuum limit.  The necessity of the decoupling
mechanism is associated with the absence of fine tuning of the curvature
and the Coulomb coupling constants.  This is a manifestation of the fact
that we have phase transition with respect to a critical line that
seperates two distinct regions [4].

 The effective action obtained after the Guassian integration of the
gauge field sector is:
$$I_{eff} = \mu_{0}\int_{0}^{1} d\lambda \sqrt{{\dot x}^{2}}+
{1\over t}\int_{0}^{1} d\lambda {\sqrt{{\dot x}^{2}
{\ddot x}^{2}-({\dot x}.{\ddot x})^{2}}
\over {\dot x}^{2}}+ {1\over 2t}\int_{0}^{1}\int_{0}^{1}
d\lambda d\lambda' {\dot x}(\lambda){\dot x}(\lambda')
V(|x-x'|)\eqno{(1 a)}$$
where the first term is the arc-length $ds=
d\lambda \sqrt{{\dot x}^{2}}$ of a point particle of bare
mass $\mu_{0}$, the second term is the curvature
k(s) = $| {d^{2}x(s)\over ds^{2}} |$  of the world line defined to
be the length of the acceleration, t is  a dimensionless coupling
constant (scale invariance of the curvature term) and V is the
long range Coulomb potential:
$$ V(|x-x'|,a)= {2g\over \pi}
{1\over |x(\lambda)-x(\lambda')|^{2}+a^{2}}\   .\eqno{(1 b)}$$
We have introduced the cut-off "a" to avoid the singularity at
$\lambda=\lambda'$ and define $ g=t.\alpha_{Coulomb}=t.{e^{2}
\over 4\pi}$.  In the arc-length gauge ${\dot x}^{2}=1$ we
can gauge fix the action and obtain:
$$I_{g.f} ={1\over 2}\mu_{0}L +{1\over 2t}\int_{0}^{L}
d\lambda(e^{-1}{\ddot x}^{2} + e + \omega({\dot x}^{2} - 1) )
         + {1\over 2t}\int_{0}^{L}\int_{0}^{L} d\lambda d\lambda'
           {\dot x}(\lambda){\dot x}(\lambda')V(|x-x'|,a)\eqno{(2)}$$
where e is an einbein to remove the square root of the acceleration,
$\omega$ is a lagrange multiplier that enforces the constraint
${\dot x}^{2}=1$, and L is the length of the path.  The partition
function is:
$$ Z =\int D\omega De Dx exp(-I_{g.f})\  .\eqno{(3)}$$

{\bf II-Large D analysis, absence of Coulomb interactions}

The effective action is obtained by integrating over
$x^{\nu}, \nu=1,...D$ we have:
$$ {S_{0}}_{eff} = {1\over 2t}\int d\lambda
e(\lambda)-\omega(\lambda) + tDtrln A\eqno{(4)}$$
where A is the operator
$$ A = \partial^{2}e^{-1}\partial^{2} -\partial\omega\partial
\  .\eqno{(5)}$$
In the large D limit the stationary point equations resulting from
varying $\omega$ and e respectively are:
$$1=tDtrG\eqno {(6 a)}$$
$$1=tDtr(e^{-2}(-\partial^{2}G))\eqno {(6 b)}$$
where the world line Green's function is defined by:
$$ G(\lambda,\lambda') = <\lambda|(-\partial^{2})A^{-1}|\lambda'>
\eqno{(7)}$$
The stationary points are:
$$ \omega(\lambda) = \omega^{*},~~~~~~~~<\lambda|e^{-1}|\lambda'>=
\int{dp\over 2\pi} {e^{i(\lambda-\lambda')}\over |p|}\eqno{(8)}$$
where $\omega^{*}$ is a constant.  Thus eqs.(6) becomes the single
mass gap equation [5]:
$$ 1=Dt\int{dp\over 2\pi}{1\over |p|+\omega^{*}}\eqno{(9)}$$
which yields
$$ \omega^{*} = \Lambda e^{-{\pi\over Dt}}\eqno{(10)}$$
where $\Lambda={1\over a}$ is an U.V. cut-off and $\omega^{*}$
is now the mass associated with the propagator:
$$ <{\dot x}^{\mu}(p){\dot x}^{\nu}(-p)>
 = Dt{\delta^{{\mu}{\nu}}\over |p|+\omega*}\eqno{(11)}$$
To obtain a  non-zero phase transition temperature the mass gap
equation must be infra-red finite for $\omega^*=0$.  Therfore without
Coulomb long range interactions the theory exists only in the
disordered phase $t>t_{c}$ and the U.V stable fixed point is $t_{c}=0$.
{}From (10) it is evident that the beta function of the pure curvature
theory is asymptotically free indicating the absence of the curvature
term at large distance scales.  In contrast to the naive classical
limit the theory is therefore well behaved and free of ghosts.
In sec. IV we will calculate the poles of the Green's
function in the presence of Coulomb interactions using Large D limit
nd show the absence of ghosts and tachyons in both the ordered and
disordered phases of the theory.
\eject

{\bf III-Phase transition in the presence of Coulomb interactions}

The integration is no longer Gaussian.  Thus we consider
$$x^{\nu}(\lambda) = x^{\nu}_{0}(\lambda) + x^{\nu}_{1}(\lambda)$$
and expand the action (2) to quadratic order in $x^{\nu}_{1}(\lambda)$
about the background straight line $x_{0}$.  The x-integration is now
Gaussian and the new effective action $S_{eff}$ is given by (4) and
(5) with the new operator A that includes the Coulomb potential
contributions.  Using the stationary solutions (8) $A_{new}$ is
given by:
$$ trln A_{new} = \int{dp\over 2\pi}ln[|p|^{3} + p^{2}\omega^* +
p^{2}V_{0}(p) + V_{1}(p)]\eqno{(12)}$$
where
$$ V_{0}(p) = {2g\over a}e^{-a|p|},~~~~~V_{1}(p)={2g\over a^{2}}
[e^{-a|p|}(|p|+ {1\over a}) -{1\over a}]\  .\eqno{(13)}$$
The new mass gap equation (6) is:
$$ 1 =Dt\int{dp\over 2\pi}{p^{2}\over p^{2}(|p|+\omega^{*}) +
p^{2}V_{0}(p) + V_{1}(p)}\eqno{(14)}$$
The critical line is defined by eq.(14) at $\omega^{*}=0$:
$$ 1 ={Dt\over \pi}\int_{0}^{1} dy{y^{2}\over y^{3}
+2g(y^{2}e^{-y}+ye^{-y}+e^{-y}-1)}\eqno{(15)}$$
where we made the change of variable y=ap and introduced the U.V
cut-off $\Lambda ={1\over a}:=\Lambda_{0}$ \footnote*{ In fact there
exist a $g^{*}$ for which any choice of $\Lambda$a=c leads to phase
transition as long as $g<g^{*}$.  We choose
$\Lambda=\Lambda_{Planck}$, therefore $c\geq 1$.} .
Notice that eq.(15) is finite
except at g=0 (absence of Coulomb interactions). The critical
curve distinguishing the two phases in the (t,g) plane is shown
in Fig.1.  The order parameter of the theory is the mass gap
equation (14) where $\omega^*$ is the parameter that
distinguishes that two phases.  In the disordered phase
$\omega^* >0$, while in the ordered phase it is straightforward
to show that $\omega^* =0$.   In the disordered phase the coupling
constants t and g are completely fixed by dimensional
transmutation in terms of the cut-off $\Lambda$ and $\omega^*$.
Thus they cannot be fine tuned.  This is an important property that
is vital in proving the absence of ghosts in our model.   From (15)
we can immediately examine whether there is a phase transition in
the pure Coulomb theory i.e ordinary QED.  The curvature term would
then be absent.  This corresponds to the absence of the $y^3$ term
in (15).  If we choose the cut-off of the theory $\Lambda$ to be at
the Compton wave length i.e $\Lambda_{Compton}={2\pi\over a}$ one
finds in this particular case that the integral (15) diverges
implying an absence of a phase transition.

{\bf IV- The Absence of Ghosts and Tachyons}

The issue of ghosts and tachyons can be examined by considering
the space-time propagator:
$$P(x_{0},y_{0}) = \int_{0}^{\infty}dL P(x_{0},y_{0},L)\eqno{(16)}$$
where
$$ P(x_{0},y_{0},L) = <\delta(x(0)-x_{0})\delta(x(L)-y_{0}>
\  .\eqno{(17)}$$
According to the Kallen-Lehmann representation there will be no ghosts
if the residue of the momentum space poles of the Fourier transform of
$P(x_{0},y_{0})$ is positive.  Since the average in (17) is with
respect to a Gaussian integral
it can be done exactly.  The result of the Fourier transform of the
propagator (16) is
$$ P(k) = \int_{0}^{\infty} dL e^{-{\mu_{0}\over 2}L +tk^{2}
G^{*}(0,L)}\eqno{(18)}$$
where
$$ G^{*}(0,L) = G(0,L)-G(0,0)\eqno{(19)}$$
and G(0,L) defined in (7) is given by:
$$ G(0,L) = {-D\over 2\pi i}\oint_{C} dp
{e^{pL}\over p^{2}(p+\omega^{*})
+ p^{2}V_{0}(p) + V_{1}(p)}\  .\eqno {(20)}$$
the curve C is a contour containing the poles of the integrand.
In the limit $a\rightarrow o$ ($\Lambda\rightarrow \infty$) we
expand the potential terms (13) in "a" neglecting terms that approach
zero as $a\rightarrow o$. Then (19) is easly evaluated we obtain:
$$G^{*}(0,L) = {1\over \beta}[-{L\over 2\Omega} + {1\over \Omega^{2}}
(1-e^{-|\Omega| L})]\eqno{(21)}$$
where
$$ \beta = 1-{4g\over 3},~~~~~~\Omega={\omega^{*}
+\Lambda g\over \beta}\  .\eqno{(22)}$$
Using (21) and (22) in (18) we finally obtain
$$ P(k) = {1\over |\Omega|}e^{\eta}\eta^{-\epsilon}
\gamma(\epsilon,\eta)\eqno{(23)}$$
where $\gamma$ is the incomplete $\gamma$ function and
$$ \epsilon={s\over |\Omega|}>0~~~~~\eta =
{Dtk^{2}\over \beta\Omega^{2}},
{}~~~~~~s={1\over 2}(\mu_{0}+{Dtk^{2}\over \beta\Omega})
\  .\eqno{(24)}$$
The poles of $\gamma$ are:
$$ \epsilon=0,~~~~~ \epsilon=-n,~~~~~~ n=1,2,....$$
using (22) these correspond to physical particles of squared masses:
$${g\over Dt}\mu_{0}\Lambda~~;~~~~~{2ng^{2}\over |\beta| Dt}
\Lambda^{2}~~;~~~~~ n=1,2,..\eqno{(25)}$$
{}From the pole structure of the $\gamma$ function, the residue of the
first particle is positive while for the rest it goes like $(-1)^{n}$.
Thus the odd n's are ghosts with mass of order $\Lambda$.  However these
ghosts decouple from the theory when we take the continuum limit for
the following reason:
Renormalization condition implies that the physical mass of our
point particle is related to the bare mass by [6]:
$$ \mu_{0} = m_{phys}^{2}a = {m_{phys}^{2}\over \Lambda}$$
Therefore our particle with positive residue has a finite mass m,
while the ghost particles have mass of order $\Lambda$.  The value
of the ghost mass cannot be reduced to a finite value because of the
absence of fine tuning of the coupling constants [4].  This is related to
the fact that we have a phase transition with two different
regions specified by $t(g)<t_{c}(g_{c})$ for the ordered phase and
$t(g)>t_{c}(g_{c})$ for the disordered phase.  In both phases we have
dimensional transmutation where the coupling constants are determined
by the cut-off $\Lambda$ so they cannot be fine tuned.  As
$\Lambda\rightarrow \infty$ these ghosts become infinitely heavy
thus decoupling from the theory.  The case when $\beta=0$ in (22)
gives only the physical particle of mass m.

In conclusion, we have a consistent generalization of QED that
includes the rigidity of the point particle with out spoiling
renormalizablity. This Rigid model of QED has two distinct phases:
I){\it a disordered phase} which at large distances is characterized
by the absence of rigidity and essentially coincides with ordinary
QED while at short distances the electron and the positron transmute
into a Majorana-Dirac fermions [1].  II){\it an ordered phase} which
is strongly coupled characterized by the presence of the curvature term
at all distance scales.  Eventhough the curvature of the world line of
the particle is a higher derivative term in the classical theory, and
naively indicates ghosts and tachyons in the mass spectrum, we show
that the regulated quantum theory, is free of both ghosts and tachyons.
This remarkable situation in the ordered phase, is ultimately related to
the fact that the phase transition implies that
the dependence on the coupling constants in the theory is
{\it non-analytic}.  Therefore we cannot analytically continue
to the {\it troubled classical theory}. The absence of ghosts and tachyons
in the disordered phase is easily understood by the fact that
the curvature coupling constant runs and approaches zero at
large distances.

{\bf Acknowledgment}

 We are very grateful to prof. A. Polyakov for his constant
encourgement and extensive support and especially suggesting
his conjecture that a  higher derivative regulated quantum
theory does not necessarily have the pathologies of the classical
theory. One of us M.A is thankful to him for long and patient
conversations explaining the decoupling mechanism of ghosts
and its connection with critical phenomena.  We are also grateful
to Prof. Y. Nambu for his extensive support, encourgement
and long discussions over the last two years with out which
we could not have gone far in our investigations.
We also Thank J. Clark and D.Seifu for the graphical representation
of the critical line and using numerical methods for locating
the poles of the Green's function.

{\bf References}

\item {[1]} M. Awada and D. Zoller, Phys.Lett B299 (1993) 151,
also see the detailed version: Cincinnati preprint Jan. 1993-105.
To appear in Int. Journal of Physics. A
\item {[2]} M.Awada, M.Ma, and D.Zoller Mod.phys. Lett A8,(1993),
2585
\item {[3]} A. Polyakov , Les Houches lecture notes 1988,
ed. E.Brezin and J.Zinn-justin.
\item {[4]} A. Polyakov, private communications.
\item {[5]} R. Pisarski, Phys. Rev. D34, 670 (1986).
\item {[6]} A. Polyakov,  Gauge fields, and Strings,
Vol.3, harwood academic publishers

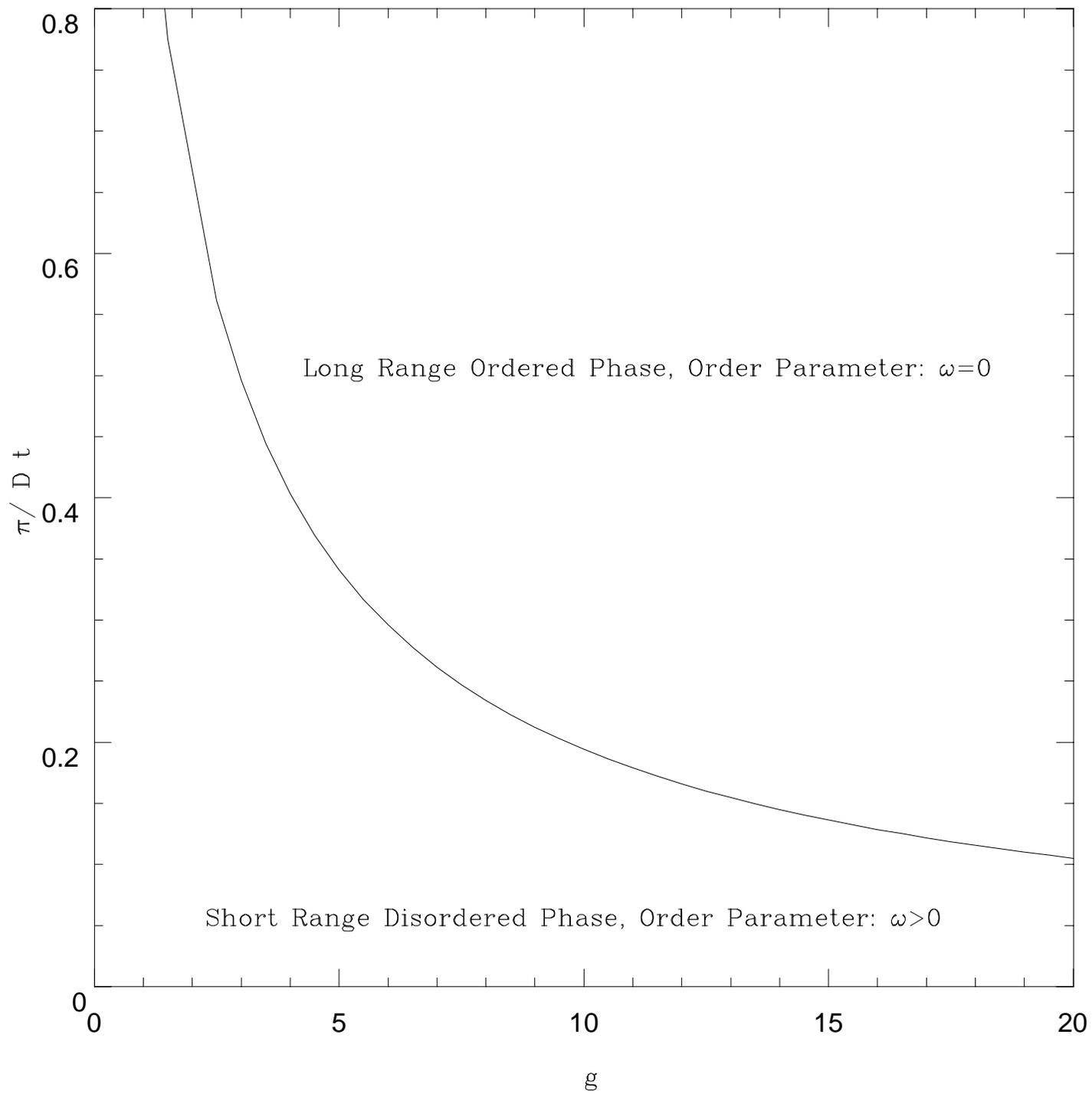
\end